\documentclass[aps,prb,showpacs,twocolumn,superscriptaddress]{revtex4-1}
\usepackage{times}
\usepackage{mathptmx}
\usepackage{amsmath}
\usepackage{txfonts}
\usepackage{bm}
\usepackage{graphicx}
\usepackage{dcolumn}
\usepackage{tabularx}
\usepackage{natbib}
\usepackage{hyperref}
\begin{document}


\title{Relationship between conductance fluctuation and weak localization in graphene}
\author{D. Terasawa}
\email{terasawa@hyo-med.ac.jp}
\author{A. Fukuda}
\affiliation{Department of Physics, Hyogo College of Medicine, Nishinomiya 663-8501, Japan}

\author{A. Fujimoto}
\affiliation{Applied Physics, Faculty of Engineering, Osaka Institute of Technology, Osaka 535-8585, Japan}

\author{Y. Ohno}
\affiliation{Graduate School of Science and Technology, Tokushima University, Tokushima 770-8501, Japan}
\author{Y. Kanai}
\author{K. Matsumoto}
\affiliation{The Institute of Scientific and Industrial Research, Osaka University, Ibaraki 567-0047, Japan}

\date{\today}

\begin{abstract}
The relationship between the universal conductance fluctuation and the weak localization effect in monolayer graphene is investigated. By comparing experimental results with the predictions of the weak localization theory for graphene, we find that the ratio of the elastic intervalley scattering time to the inelastic dephasing time varies in accordance with the conductance fluctuation; this is a clear evidence connecting the universal conductance fluctuation with the weak localization effect. We also find a  series of scattering lengths that are related to the phase shifts caused by magnetic flux by Fourier analysis. 
\end{abstract}




\maketitle


\section{\label{intro}Introduction}

Recent progress in condensed matter physics has enabled researchers to study the novel concepts that occasionally accompany exotic particles\,\cite{NovoselovGeim,PhilipKim,Hasan,Qi,Willett}.
These particles are detectable using matter-wave properties, thus quantum interference (QI) effects have become increasingly important in research.
These interference effects are evident in the quantum transport phenomena, therefore, QI corrections to the conductivity have been the subject of study for many decades\,\citep{AndersonLocalization,Bergmann,LeeStone,*LeeStoneFukuyama}.
Recent investigations of graphene, the ``playground'' of chiral Dirac fermions, have revealed its interesting QI corrections, as compared with conventional electron gases in metals and semiconductors in two well-known examples\,\cite{Ojeda-Aristizabal,Bohra,SuzuuraAndo,Morpurgo,McCann,Morozov,Tikhonenko,Tikhonenko2,Iagallo,Lundeberg,
AMRBaker}.
The first case is the failure of the universality of the conductance fluctuation\,\cite{Ojeda-Aristizabal,Bohra}, since the order of the fluctuation amplitudes between the variations in the Fermi energy and the magnetic field are different.
The second case is the essential contribution of elastic scatterings to the weak localization (WL) effect\,\cite{SuzuuraAndo,Morpurgo,McCann,Morozov,Tikhonenko,Tikhonenko2} due to the properties of a chiral two-Dirac cone system.
Furthermore, weak electron-phonon scattering effects\,\cite{Tikhonenko2,Stauber,Hwang} and the spin-orbit interaction\,\cite{Huertas-Hernando} render the WL in graphene unique, even when compared to the WL effect in the other Dirac fermion system that exists on topological insulator surfaces\,\cite{MinhaoLiu} or another two-dimensional material\,\cite{Du}.
Therefore, it is important to investigate these interference effects in graphene in particular, as it is a stimulating material that can be used to extend the boundaries of two-dimensional materials research\,\cite{Novoselov26072005,Lego}.

Thus far,  a number of papers that discuss the relationship between universal conductance fluctuation (UCF) and WL both experimentally and theoretically have been published\,\cite{Horsell,KechedzhiUCF,Chuang,Westervelt1,Westervelt2,YungFuChen,Borunda,Minke}.
In particular, double experimental papers\,\cite{Westervelt1,Westervelt2} using scanning probe microscopy for graphene and transport experiments\,\cite{Tikhonenko,YungFuChen} revealed that the dephasing length causing the UCF and WL are of the same order.
However, analysis of the data is usually undertaken in order to separate the effects of UCF from the WL.
Therefore, no direct evidence connecting WL and UCF has ever been proposed.

In this paper, we demonstrate that the origin of the UCF in graphene can indeed be attributed to the WL effect.
As mentioned earlier, the WL effect is caused by elastic scattering as well as inelastic scattering, and importantly, these scattering times can be deduced through the analysis of the WL effect.
We find that the ratio of the inelastic dephasing time to the elastic intervalley scattering time varies in accordance with the zero-field resistance fluctuation.
Furthermore, our fast Fourier transform (FFT) analysis of the fluctuation as a function of the magnetic field also reveals 
an effective length scale that is related to the Aharonov-Bohm (AB) effect\,\cite{ABeffect}.
We discuss the relationship between this effective length and the intervalley scattering length.

The remainder of this paper is structured as follows.
In Section \ref{experiment}, the experimental methods and procedures are described.
In Section \ref{results}, we present our experimental results and discussion, beginning by examining the WL behavior and the relationship between the UCF and WL (\ref{VgDependence}) ; followed by an analysis and discussion on the magnetic field dependence of the UCF (\ref{BDependence}). 
Finally, we present our summary in Section \ref{summary}.

\section{\label{experiment}Experimental details}

Three graphene samples, peeled from kish graphite, are fabricated using a mechanical exfoliation method \cite{Novoselov_science} on an SiO$_2$ surface that is separated by 300\,nm from an $n^+$-doped Si substrate.
We chose three monolayer flakes of graphene, A, B and C, and the contact patterns were fabricated using electron beam lithography (see the right panels of Fig.\,\ref{fig1}).
Two of the samples (A and B) are shaped into Hall bar patterns and the third sample (C) is contacted at opposite edges of the sample. 
Ohmic contact materials (10-nm-thick Pd followed by 100-nm-thick Au) were deposited through thermal evaporation, followed by a liftoff process in warm 1-methyl-2-pyrrolidone.
The number of layers was confirmed by analysis of the Raman shift spectra\,\cite{Ferrari,Ferrari200747,SuppleMat}.
The samples were first annealed at 700\,K in the H$_2$ atmosphere for 30 min,
then they were placed in a sample cell containing a resistance heater one-by-one.
Then the samples were annealed again {\it in situ}, using the heater  
for 2\,h at $\sim 410$\,K before cooling.
This allows us to desorb molecules on the graphene surface, and shifts the charge neutral point (CNP)  to approximately 0\,V.
The resistances are measured using a standard ac lock-in technique with a frequency of 37\,Hz.
The carrier densities are controlled by the back gate voltage, $V_{\rm g}$, which is applied between the graphene sheet and the substrate, in accordance with the relation $dn_{\rm d}/dV_{\rm g} = 7.2 \times 10^{10}$\,cm$^{-2}$V$^{-1}$.
A He-free Gifford-MacMahone refrigerator is used to cool both the sample cell (and the contained sample) and a superconductor magnet with a maximum magnetic field of 8\,T.

\section{\label{results}Results and discussion}
\subsection{\label{VgDependence}Relationship between UCF and WL}

\begin{figure}  
\includegraphics[width=0.86\linewidth]{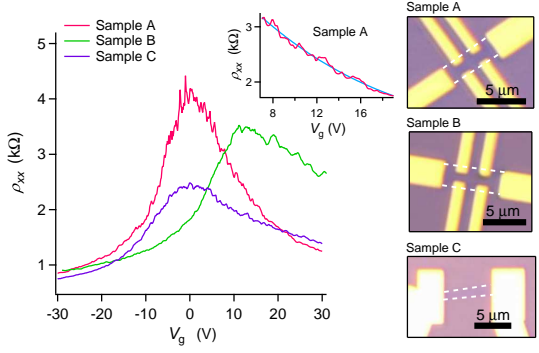}
\caption{\label{fig1}(Color online)\,$\rho_{xx}$ of sample A, B, and C as a function of $V_{\rm g}$ at 4.4\,K (except for C at 5.7\,K) and $B=0$\,T. Inset: Expanded view of $\rho_{xx}$ for the electron side of sample A. Right panels:\,Optical image of the examined graphene devices, sample A, B, and C. }
\end{figure}

Figure \ref{fig1} shows the longitudinal resistivity $\rho_{xx}$ of samples A, B, and C as a function of $V_{\rm g}$ at a temperature of 4.4\,K (except for C at 5.7\,K) and at a magnetic field $B$ of 0\,T.
For the sample A, the value of the gate voltage at the CNP, $V_{\rm CNP}$, at which the carrier type changes from electrons ($V_{\rm g} > V_{\rm CNP}$) to holes ($V_{\rm g} < V_{\rm CNP}$), is approximately $ -1$\,V.
As we can see, $\rho_{xx}$ exhibits asymmetric dependence on $V_{\rm g}$, owing to the invasive contacts with Pd\,\cite{Huard}. Therefore, the value of the carrier mobility, $\mu$, differs in both cases, with $\mu =1/(n_{\rm d}e\rho_{xx}) \approx 0.25$ and $0.40$\,m$^2$V$^{-1}$s$^{-1}$ for electrons and holes, respectively, at $|V_{\rm g} -V_{\rm CNP}| \simeq 20$\,V.
For $-30 \leq V_{\rm g} \leq 30$\,V, we find $k_{\rm F} \ell \geq 5$, where $k_{\rm F}$ is the Fermi wave number and  $\ell = h/(2e^2 k_{\rm F} \rho_{xx})$ is the mean free path, indicating that the system is in a diffusive metal. 
The mobilities of samples B and C at the hole side are $\approx 0.36$ and $\approx 0.45$\,m$^2$V$^{-1}$s$^{-1}$, respectively.
Hereafter, we feature the results of sample A because we have obtained the substantially similar results for samples B and C.

\begin{figure}  
\includegraphics[width=1\linewidth]{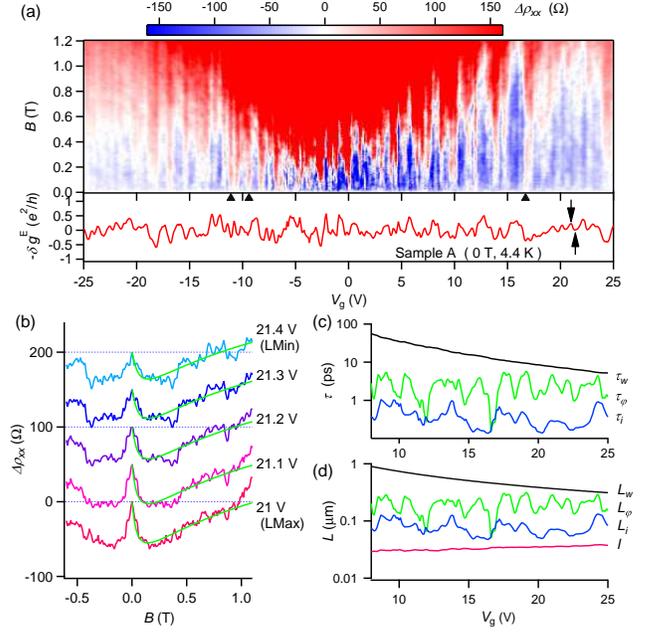}
\caption{\label{fig2}(Color online) (a)\,(Upper Panel) Image plot of ${\it \Delta}\rho_{xx}$ as a function of $B$ and $V_{\rm g}$. (Lower Panel) $-\delta \varg^{\rm E}$ as a function of $V_{\rm g}$. The small black triangles indicate the positions at WAL.%
(b)\,Typical fitting results from LMax to LMin (indicated by the arrows in the lower panel of (a)) for electron region (each trace is offset for $+50$\,$\Omega$).
(c)\,$\tau_\varphi$ and $\tau_i$ in the log scale obtained from fitting as a function of $V_{\rm g}$ along with $\tau_w$. (d)\,$L_\varphi$ and $L_i$ converted from $\tau_\varphi$ and $\tau_i$ in (c) along with $\ell$ and $L_w$ in the log scale as a function of  $V_{\rm g}$. Note that $\tau_w$ is calculated as  $\tau_w^{-1} = 2\tau_0(\eta \epsilon_{\rm F}^2/(\hbar v_{\rm F}^2))^2$, where $\tau_0 = \ell/(2v_{\rm F})$ denotes the momentum relaxation time, $\epsilon_{\rm F} =\hbar k_{\rm F} v_{\rm F}$ denotes the Fermi energy, and $\eta$ denotes the structural parameter equal to $\eta = \gamma_0a^2/(8\hbar^2)$, with $\gamma_0 \approx 3$\,eV (the nearest-neighbor-hopping energy) and $a \approx 0.26$\,nm (the lattice constant in graphene), and $L_w = \sqrt{D\tau_w}$.}
\end{figure}

Then we measure the magnetic field, $B$, dependence.
Figure \ref{fig2} (a) shows the image plot of the changes in the resistivity, ${\it \Delta} \rho_{xx}$,  as a function of $B$ and $V_{\rm g}$, i.e., 
${\it \Delta}\rho_{xx}(B,V_{\rm g}) = \rho_{xx}(B,V_{\rm g}) - \rho_{xx}(0,V_{\rm g})$,  for $0 \leq B  \leq 1.2$\,T and $-25 \leq V_{\rm g} \leq 25$\,V at 4.4\,K.
This sample shows an asymmetric dependence under the inversion of $B$; therefore, we employ ${\it \Delta}\rho_{xx}(B) = [{\it \Delta}\rho_{xx}(+B) + {\it \Delta}\rho_{xx}(-B)]/2$ (see e.g.\,\cite{Jouault} and \cite{SuppleMat}).
We label $\rho_{xx}(0, V_{\rm g})$ at 4.4\,K as $\rho_{xx}^0$.
For comparison, we plot the conductance fluctuation, $- \delta \varg^{\rm E} = \delta\rho_{xx}^{\rm E}/\langle \rho_{xx}^0 \rangle^2 = (\rho_{xx}^0 - \langle \rho_{xx}^0 \rangle)/\langle \rho_{xx}^0 \rangle^2$, in the lower panel of Fig.\,\ref{fig2} (a).
We obtain $\langle \rho_{xx}^0 \rangle$ by polynomial fitting $\rho_{xx}^0$ (the blue line of Fig.\,\ref{fig1} inset).
The white regions correspond to ${\it \Delta}\rho_{xx} \sim 0$, the blue regions to ${\it \Delta}\rho_{xx} <0$, and the red regions to ${\it \Delta}\rho_{xx} >0$.
In this sample, negative magnetoresistance (${\it \Delta}\rho_{xx} < 0$) is observed at small $B$ for the measured $V_{\rm g}$ region, which is a clear indication of the suppression of the WL by the magnetic field.
However, positive magnetoresistance ${\it \Delta} \rho_{xx} \gtrsim 0$ is observed in some conditions, for example $V_{\rm g} \approx -11$ and 17\,V, for which the system shows weak anti-localization (WAL).
The blue region is extended for higher fields at the local maximum (LMax) of $\rho_{xx}^0$.
In contrast, the blue region ends at lower fields at the local minimum (LMin) of $\rho_{xx}^0$.
In general, the blue region is extended for higher fields as the value of $V_{\rm g}$ deviates from the CNP.

The magnetic field correction of the resistivity in graphene due to the WL effect has been analyzed by McCann {\it et al}.\,\cite{McCann}.
Owing to Berry's phase of the chiral carriers, carriers within one valley in the graphene momentum space cannot be perfectly backscattered\,\cite{Ando}. 
This results in positive conductance corrections (WAL)\,\cite{SuzuuraAndo}.
However, defects on the atomic scale can induce intervalley scattering events, and satisfy the constructive interference of a closed scattering path.
As a result of their analysis of the WL in the presence of $B$, they proposed the following equation for ${\it \Delta} \rho_{xx}$:
\begin{gather}
{\it \Delta} \rho_{xx}^{\rm th}(B) = -\frac{e^2 {\rho^0_{xx}}^2}{\pi h} \left [ F \left(\frac{B}{B_\varphi}\right)-F\left(\frac{B}{B_\varphi+2B_i}\right) - 2F\left( \frac{B}{B_\varphi + B_\ast} \right) \right ],  \label{eqDR} \\
F(z) = \ln z+\varPsi \left( \frac{1}{2} + \frac{1}{z} \right),\mspace{18mu} B_{\varphi, i, \ast} = \frac{\hbar}{4De}\tau^{-1}_{\varphi, i, \ast}. \label{eqtau}
\end{gather}
Here,  $\varPsi$ is the digamma function, $D$ denotes the diffusion constant obtained from $\ell$ as $D= v_{\rm F}\ell/2$ with $v_{\rm F} \approx 1.0 \times 10^6$\,m/s (Fermi velocity),
$\tau_\varphi$ and $\tau_i$ represent the dephasing time and intervalley scattering time, respectively.
$\tau_\ast^{-1} \equiv \tau_i ^{-1} + \tau_z^{-1} + \tau_w^{-1}$, where $\tau_z$ denotes the intravalley scattering time, and $\tau_w$ denotes the intravalley warping time combined with the chirality breaking time\,\cite{Morpurgo}.
According to Eq. (\ref{eqDR}), the WL magnetoresistance is realized for $B_{\varphi} < B_i$ ($\tau_\varphi > \tau_i$). 
The first term represents the carrier dephasing and determines the resistance curvature for $B < B_i$.
The second and third terms contribute to the antilocalization, and the theory estimates $B_\ast \sim B_i$ ($\tau_i \ll \tau_z, \tau_w$) for the WL in the electron and hole regions (but not for a high-carrier-density region, where the warping effect becomes important). 

We fit the data shown in Fig.\,\ref{fig2}\,(a) using Eq.\,(\ref{eqDR}).
Since $B_i$ and $B_\ast (B_z)$ are the same contribution in Eq.\,(\ref{eqDR}), we cannot uniquely determine $B_i$ and $B_\ast (B_z)$ at the same time. 
Therefore, we preferentially decide $B_i$ in order to satisfy the above conditions ($B_i > B_\varphi, B_i > B_z$).
Typical fitting results from LMax to LMin are displayed in Fig.\,\ref{fig2}\,(b).
We fit the data point-by-point for $V_{\rm g}$ (0.1-V step) in the $0 < B < 0.8$\,T range. Good agreements for $0 \leq B \leq B_{\rm tr}  (\simeq 0.54$\,T at $V_{\rm g} =20$\,V), where the diffusive theory of WL is applicable, are obtained for the electron and hole regions without the inclusion of a fitting parameter before the square bracket in Eq.\,(\ref{eqDR}).
Here, $B_{\rm tr}$ denotes the transport magnetic field, which is obtained based on the condition that the magnetic length is approximately equal to $\ell$ ($B_{\rm tr} = \hbar /(e\ell^2)$ with $\hbar = h/(2\pi)$). 
However, the fitting is poor near the CNP.
This may be due to the electron-hole paddles\,\cite{Martin} formed near the CNP, in which Eq.\,(\ref{eqDR}) is not applicable. 
We show the fitting results of $\tau_{\varphi}$ and $\tau_i$ in Fig. \ref{fig2} (c) and the scattering lengths $L_{\varphi}$ and $L_i$ converted from the corresponding scattering times  in Fig. \ref{fig2} (d) for the electron region as $L_{\varphi,i} = (D\tau_{\varphi,i})^{1/2}$.
Since the system shows the WL, $L_{\varphi} > L_i$ for almost the entire region; however, there exist some points that show $L_{\varphi} \lesssim L_i$ (WAL)\,\cite{Tikhonenko2}.

\begin{figure}  
\includegraphics[width=1\linewidth]{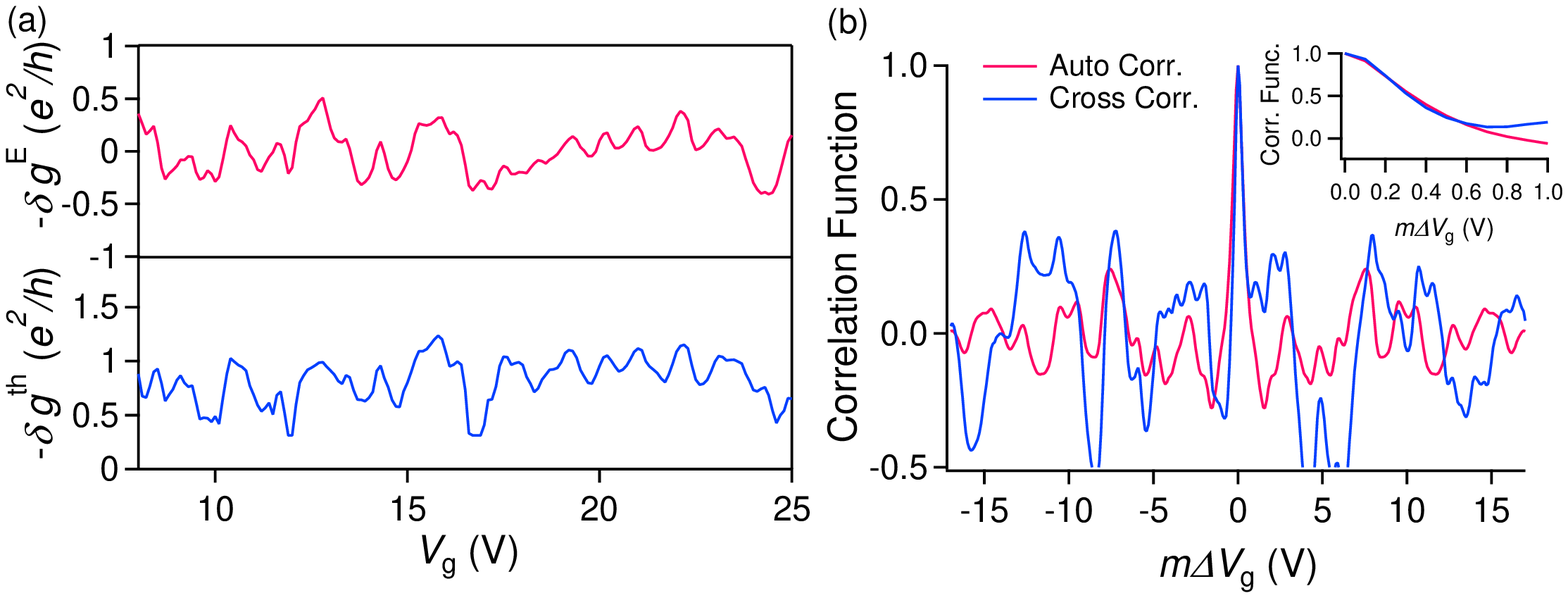}
\caption{\label{fig3}(Color online) (a)\,Comparison of $ -\delta \varg^{\rm E}$ and $-\delta \varg^{\rm th}$ obtained from the fitting (in units of $e^2/h$) as a function of $V_{\rm g}$. (b)\,Normalized auto-correlation of $ -\delta \varg^{\rm E}$ and cross-correlation between  $ -\delta \varg^{\rm E}$ and $ -\delta \varg^{\rm th}$. }
\end{figure}

Furthermore, we consider another equation in McCann {\it et al}.\,\cite{McCann}. 
According to their WL analysis, theoretical changes in resistivity $\delta \rho_{xx}^{\rm th}$ at 0\,T depend on the $\tau_\varphi / \tau_i$ ratio, as \cite{note_amp}
\begin{equation}
\frac{\delta \rho_{xx}^{\rm th}}{\langle \rho_{xx}^0 \rangle} = -\delta \varg^{\rm th} \approx \frac{e^2}{\pi h} \cdot \ln \left [ 1+2\frac{\tau_\varphi}{\tau_i} \right ]. \label{eqRTau}
\end{equation}
The lower panel of Figure \ref{fig3} (a) shows $-\delta \varg^{\rm th} $ with obtained values of $\tau_\varphi / \tau_i$ from the WL fitting. 
As we can see, $-\delta \varg^{\rm th}$ varies as if it resembles the behavior of $-\delta \varg^{\rm E}$.
Their similarity becomes evident when we compare the normalized auto-correlation function of $-\delta \varg^{\rm E}$, 
${\mathcal F}_{\rm auto}(m\Delta V_{\rm g}) =\sum_{m=-\infty}^{+\infty} \delta \varg^{\rm E}(V_{\rm g})\cdot \delta \varg^{\rm E}(V_{\rm g} +m\Delta V_{\rm g})/(\delta \varg^{\rm E}(V_{\rm g})\cdot \delta \varg^{\rm E}(V_{\rm g})) $, 
and the cross-correlation function between $-\delta \varg^{\rm E}$ and $-\delta \varg^{\rm th}$, ${\mathcal F}_{\rm cross}(m\Delta V_{\rm g}) =\sum_{m=-\infty}^{+\infty}  \delta \varg^{\rm E}(V_{\rm g})\cdot \delta \varg^{\rm th}(V_{\rm g} +m\Delta V_{\rm g})/(\delta \varg^{\rm E}(V_{\rm g})\cdot \delta \varg^{\rm th}(V_{\rm g}))  $, where $m= \cdots, -2,-1,0,1,\cdots$, $\Delta V_{\rm g}=0.1$\,V, 
with fitting values of $\tau_\varphi$ and $\tau_i$  (Fig.\, \ref{fig3} (b)).
Both functions show a striking resemblance, since the full width at half maximum of the cross-correlation function shows an equivalent value to that of the autocorrelation and no other strong peaks are observed in the cross-correlation, showing that the unique $-\delta \varg^{\rm E}$ fluctuation is reproduced by Eq.\,(\ref{eqRTau}) .
Therefore, we conclude that the UCF in graphene that is induced by the $V_{\rm g}$ sweep can be attributed to the variation in $\tau_\varphi/\tau_i$, and consequently the UCF can be described by the WL equations (\ref{eqDR}) and (\ref{eqRTau}).

It is believed that the long-range Coulomb interaction cannot break the chirality of the carriers. Atomically sharp defects, such as vacancies, dislocations and corrugations, have a very short-ranged effect and, therefore, intervalley scattering can occur\,\cite{Morozov,Tikhonenko,Tikhonenko2,Ni}.
As we can see from the previous experiment\,\cite{Tikhonenko} and the AFM data of one of our samples\,\cite{SuppleMat}, corrugation in the graphene sample enhances the elastic scattering rate, which contributes to the WL. 
Furthermore, it was shown that exfoliated graphene on a SiO$_2$ surface usually contains a non-negligible density of atomic scale defects\,\cite{Ni,Ishigami}.
We suggest that these defects in the real space affect the quantum transport trajectory of the carriers on the Fermi surface significantly, in that carriers on different Fermi surfaces take different closed paths. 
Therefore, small modulations in the Fermi surface that are induced by $V_{\rm g}$ cause significant variations in the carrier trajectory and its chirality, resulting in the UCF.
Unlike the graphene case, elastic scattering effects do not appear in WL analyses conducted on the conventional diffusive systems. 
We surmise that this peculiarity of graphene elucidates the connection between the WL and the UCF.

\begin{figure}  
\includegraphics[width=0.7\linewidth]{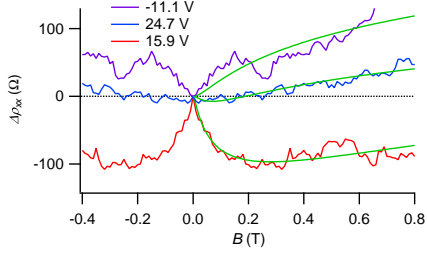}
\caption{\label{fig3half}(Color online) ${\it \Delta} \rho_{xx}$ as a function of $B$ for a typical WL case with a strong warping effect ($V_{\rm g}=15.9$\,V), a WAL case ($V_{\rm g} = -11.1$\,V), and a case that shows neither WL nor WAL ($V_{\rm g}=24.7$\,V). }
\end{figure}

Corrugations also induce lattice warpings, which contribute to the WL.
We estimate $\tau_i < \tau_w$ because the atomic scale defects should to be included in the nanometer scales, in contrast to the $\mu$m scale corrugations observed in the AFM data\,\cite{SuppleMat}. 
For prominent lattice warpings, $\tau_w$ could be less than $\tau_i$ ($\tau_w < \tau_i < \tau_\varphi$), and the theory\,\cite{McCann} predicts saturated resistance at $B \sim B_i$.
Although our fitting cannot precisely determine the magnitude of the relationship between $\tau_w$ and $\tau_i$ because both terms have the same contribution in Eq.\,(\ref{eqDR}),
perhaps the extended blue regions in Fig.\,\ref{fig2} (a), which are often observed at LMax of $\rho^0_{xx}$, correspond to this situation.
Furthermore, when $\tau_i$ becomes longer than $\tau_\varphi$, the resistance is expected to exhibit  WAL behavior, which is also observed for some $V_{\rm g}$ points.
Figure \ref{fig3half} shows three examples of magnetoresistance: saturated resistance behavior, WAL behavior, and behavior that has neither WL nor WAL ($\tau_i \sim \tau_\varphi$).
The difference between this study and the previous study\,\cite{Tikhonenko}  is worth discussing, particulary on the roles of atomic defects and lattice warping in the WL and WAL effects.
This may be attributed to the defferent sample qualities, as the samples in this study have lower mobilities and high density of atomic defects, with the latter revealed by the visible {\it D} band ($\approx 1320$\,cm$^{-1}$) in the Raman spectrum\,\cite{SuppleMat}.
A recent Raman study\,\cite{JinpyoHong} reveals that annealed graphene contains amorphous carbon generated from residual hydrocarbons at elevated temperature. 
Therefore, the annealing processes possibly further induced atomic defects in our samples.
In addition, it is interesting to compare our results with studies of induced atomic defects from vacancies introduced by ozone exposure\,\cite{Moser}, hydrogenation\,\cite{Matis}, and fluorination\,\cite{HongX}.
Our results are consistent with the referenced studies, indicating the elastic scattering length ($\sim 6$\,nm) is well correlated with the domain size of the induced defects density, which is typically one order of magnitude smaller than our elastic length $\ell$ obtained by Drude conductivity (the mean free path).

\subsection{\label{BDependence}Magnetic field dependence of UCF}

\begin{figure}  [b]
\includegraphics[width=1\linewidth]{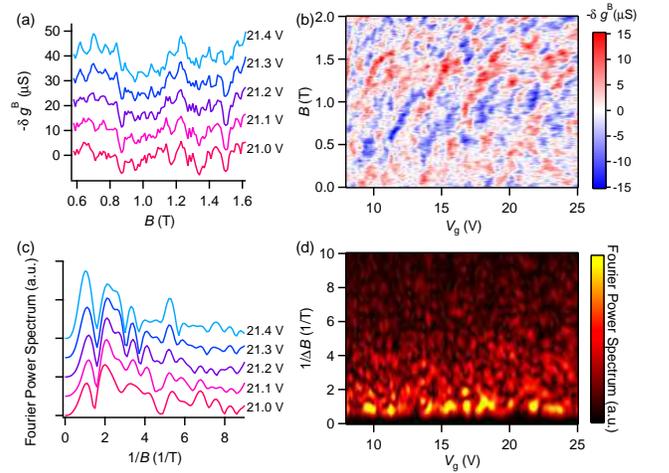}
\caption{\label{fig4}(Color online) (a)\,$-\delta \varg^{\rm B}$ as a function of $B$ for $V_{\rm g} = 21$ to 21.4 V. (b)\,Image plot of $- \delta \varg^{\rm B}$ as a function of $V_{\rm g}$ and $B$. (c)\, Fourier power spectrum of the fluctuation data in (a).  (d)\,Image plot of Fourier power spectrum of the data in (b) as a function of $V_{\rm g}$ and $1/\Delta B$.}
\end{figure}

The magnetic field dependence of the fluctuation further reveals an intriguing physics.
Figure\,\ref{fig4}\,(a) shows plots of resistance fluctuation as a function of $B$ for several $V_{\rm g}$ values
obtained by subtracting the WL fitting results $\rho_{xx}^{\rm WL}$ from ${\it \Delta}\rho_{xx}$.
However, there still remains the quadratic $B$ dependence originated from the electron-electron interaction $\rho_{xx}^{\rm EEI}$\,\cite{Jobst,Jouault}. 
Thus the conductance fluctuation as a function of $B$, $-\delta \varg^{\rm B}$, becomes $- \delta \varg^{\rm B} = \delta \rho_{xx}^{\rm B}/\langle \rho_{xx}^0 \rangle^2 = ({\it \Delta}\rho_{xx} - \rho_{xx}^{\rm WL} - \rho_{xx}^{\rm EEI})/\langle \rho_{xx}^0 \rangle^2$.
Continuous changes are observed in $-\delta\varg^{\rm B}$ from $V_{\rm g} = 21.0$ (LMax) to 21.4\,V (LMin). 
Figure \ref{fig4} (b) shows the image plot of  $- \delta \varg^{\rm B} $ as a function of $V_{\rm g}$ and $B$. 
Although complicated, we observe fluctuation patterns that vary by changing $V_{\rm g}$. 
Thus, the fluctuation is not completely random, but has some orders as a function of $V_{\rm g}$.
We display in Fig.\,\ref{fig4}\,(c) the power spectrum of the FFT of the data shown in Fig.\,\ref{fig4}\,(a).
For the FFT analysis, subtracting the background variations, i.e. the WL and EEI effects, is important because the background variations produce Fourier peaks at $1/\Delta B \simeq 0$.
Several peaks are obtained in the Fourier spectrum, for which the peak intensities and positions gradually vary as $V_{\rm g}$ varies.
Then, we construct the FFT from the data in (b), and the results (the Fourier power spectrum) are shown in (d) as an image plot. 
This image shows a systematical change in the peak intensities and positions as a function of $V_{\rm g}$. 

\begin{figure}  
\includegraphics[width=1\linewidth]{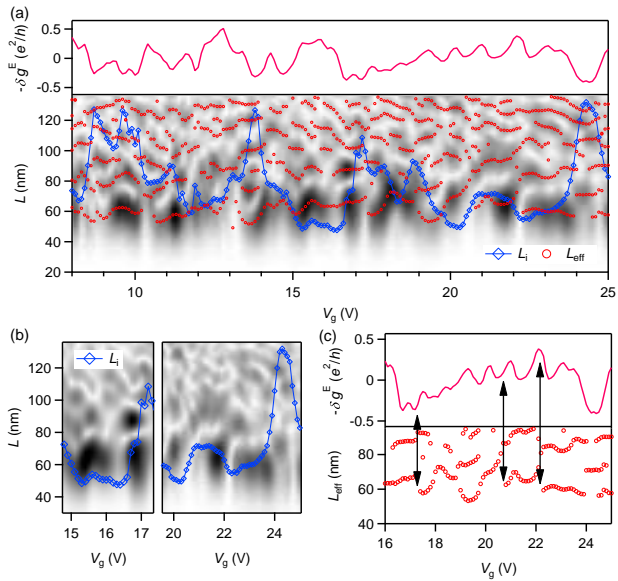}
\caption{\label{fig5}(Color online) (a)\,Comparison of $-\delta \varg^{\rm E}$,  $L_{\rm eff}$, and $L_i$ as a function of $V_{\rm g}$. The background image plot represents the FFT power spectrum (same as Fig.\,\ref{fig4}\,(d) ) with the vertical axis converted to $L = (\phi_0 \cdot \Delta (1/B))^{1/2}$. The darker the black colors represent higher spectrum peaks. (b)\,Detailed comparison between FFT spectrum image and $L_i$ for $14.8 \le V_{\rm g} \le 17.3$\,V (left panel) and $19.6 \le V_{\rm g} \le 25.0$\,V (right panel). (c)\,Detailed comparison between $-\delta \varg^{\rm E}$ and $L_{\rm eff}$. The both sides arrows indicate typical points between disconnections in $L_{\rm eff}$ and corresponding LMax and LMin of $-\delta \varg^{\rm E}$.}
\end{figure}

Taking into account the AB effect\,\cite{ABeffect}, the UCF occurs as a consequence of phase shifts by a magnetic flux, thus we suppose that the carriers encircle a magnetic flux on the scale of effective areas $A_{\rm eff}$. Then the $n$-th peak position value $\Delta(1/B)^{(n)}_{\rm peak}$ is  converted into the  $n$-th effective area $A_{\rm eff}^{(n)} = \phi_0 \cdot\Delta(1/B)^{(n)}_{\rm peak} (n=1,2,3, \cdots)$, which immediately leads to the effective length $L_{\rm eff}^{(n)} = \bigl(A_{\rm eff}^{(n)}\bigr)^{1/2}$.
Here, $\phi_0 = h/e$ denotes the magnetic flux quantum.
Consequently, we obtain several series of $L_{\rm eff}$ corresponding to each FFT peak.

Figure \ref{fig5} (a) shows $L_{\rm eff}$ as a function of $V_{\rm g}$ along with $L_i$. Although $L_i$ varies from 40 to 130\,nm, we notice that the shortest length series of $L_{\rm eff}$ and lower $L_i$ lengths are on the same scale (see also Table\,\ref{tab1}).
Furthermore, we observe that $L_i$ and $L_{\rm eff}$ have close values at many points,
and for many regions, $L_i$ varies as if it follows the contour of the leading edge of peaks, for example at $15 \leq V_{\rm g} \leq 16$\, V and  around $V_{\rm g} \approx 20$\, V (Fig.\,\ref{fig5} (b)).
Also, the relationship between $-\delta \varg^{\rm E}$ and $L_{\rm eff}$ is interesting,
 in that $L_{\rm eff}$ breaks up or disconnects suddenly, often at LMax or LMin of $-\delta \varg^{\rm E}$ (Fig.\,\ref{fig5} (c)), suggesting the beginning of a new series of $L_{\rm eff}$.
This result leads us to a model of the carriers acquiring the AB phases for each closed path, and the fluctuation occurs owing to the variation of the QI corrections from the AB phases.
Among the closed paths, the one that includes the intervalley scattering corresponds to $L_i$, and contributes to WL.
The phase acquired by a carrier becomes complicated owing to the chiral Berry's phase and the AB phase,
and it should be different if a closed path includes an intervalley scattering event or not.
Furthermore, atomic scale defects in graphene can produce fictitious gauge fields that cause AB interferences\,\cite{juan2011aharonov-bohm}.
Taking into account that atomic scale defects cause the intervalley scattering, this further corroborates the obtained length scale matches between $L_i$ and $L_{\rm eff}$. 
A number of previous experimental data show fluctuations in $\rho_{xx}$ as a function of $B$\,\cite{Tikhonenko,Jouault,Chuang,YungFuChen,Pezzini}. 
We believe that we can extract $L_{\rm eff}$ from these fluctuations by proper analysis.


\begin{table}
\caption{\label{tab1}\,Comparison of typical scattering lengths (nm) and $\tau_\varphi/\tau_i$. }
 \begin{ruledtabular}
\begin{tabular}{l|rrrrr} 
$V_{\rm g}$ &  $L_\varphi$  & $L_i$ & $\tau_\varphi/\tau_i$ & $\ell$ &   $L_{\rm eff}^{(1)}$   \\ \hline
 21.0\,V (LMax) &  290  & 71 & 16 & 35 & 66 \\ \hline
 21.4\,V (LMin) &  210  & 72 & 8.5 & 36 & 65 \\ 
\end{tabular}
\end{ruledtabular}
\end{table}

Additionally, we wish to point out that the fluctuation amplitude of the $B$ sweep (e.g. at $V_{\rm g} =21$\,V is $ \sim 10\,{\rm \mu S} \sim 0.25 \cdot(e^2/h)$) is smaller than the variations observed with regard to the gate voltage dependence ($\sim 0.5 \cdot(e^2/h)$, see Fig.\,\ref{fig2}\,(a) lower panel); this is a clear contradiction of the ergodic hypothesis of the UCF\,\cite{LeeStoneFukuyama}, that has also been observed in other experiments\,\cite{Ojeda-Aristizabal,Bohra}.
This difference is attributed to the main finding of this experiment; i.e., that the UCF in graphene is related to the WL effect.
This suggests that the UCF in graphene is directly related to the time-reversal symmetry of the closed trajectories, which is broken in the presence of $B$.
Furthermore, this is consistent with the simulation results of Rycerz {\it et al}.\,\cite{Rycerz}, implying that the UCF in graphene is caused by variations in trajectories rather than phase shifts.

\section{\label{summary}Summary}

In summary, we have measured the effects of the QI correction on electric transport phenomena in monolayer graphene.
We have clarified that the UCF in graphene can indeed be attributed to the  WL effect.
We have also observed that the conductance fluctuation as a function of $B$ is caused by QI corrections of the AB effect through the FFT analysis. 
We believe that the analysis conducted in this study will contribute significantly to other QI correction analyses regarding UCF.

\begin{acknowledgments}
We are grateful to H. Suzuura for a fruitful discussion, to A. Sawada and H. Yayama for advice on the He-free refrigerator experimental setups, to T. Terashima and Y. Sasaki for permission to use the clean-room facilities in the LTM center of Kyoto University, and to T. Nakajima for fabricating the cryostat and electronic parts.
This work was supported by the KAKENHI on Innovative Areas (No.\,25103722,\,``Topological Quantum Phenomena"), the JSPS KAKENHI (Nos.\,24540331, 25870966, and 15K05135), Grants-in-Aid for Researchers from Hyogo College of Medicine  2011 (A.F.) and 2015 (D.T.), and a Grant for Basic Science Research Projects from the Sumitomo Foundation.
This work was performed under the Cooperative Research Program of the ``Network Joint Research Center for Materials and Devices" (2011A10 and 2013A18).
\end{acknowledgments}


%

\end{document}